\documentclass[12pt]{article}
\usepackage{styles/mystyle} 

\begin{document}

\title{Clustering doc2vec output for topic-dimensionality reduction: A MITRE ATT\&CK calibration}

\date{\today}

\author[1]{Nathan Monnet\thanks{\href{mailto:nathan.monnet@armasuisse.ch}{nathan.monnet@armasuisse.ch}}}
\author[2]{Loïc Maréchal\thanks{\href{mailto:loic.marechal@unil.ch}{loic.marechal@unil.ch}}}
\author[1]{Julian Jang-Jaccard\thanks{\href{julian.jang-jaccard@armasuisse.ch}{julian.jang-jaccard@armasuisse.ch}}}
\author[1]{Alain Mermoud\thanks{\href{mailto:mermouda@ethz.ch}{mermouda@ethz.ch}\newline\newline This document results from a research project funded by the Cyber-Defence Campus, armasuisse Science and Technology. We appreciate helpful comments from seminar participants at the Cyber Alp Retreat 2024. Corresponding author: Loïc Maréchal e-mail: \href{mailto:loic.marechal@unil.ch}{loic.marechal@unil.ch}}}

\affil[1]{\small Cyber-Defence Campus - armasuisse, Science and Technology -- 1015 Lausanne, Switzerland}
\affil[2]{\small Department of Information Systems, University of Lausanne -- 1015, Lausanne, Switzerland}


\renewcommand{\thefootnote}{\fnsymbol{footnote}}
\singlespacing
\maketitle
\vspace{-.2in}

\begin{abstract}
We introduce a novel approach to text classification by combining doc2vec embeddings with advanced clustering techniques to improve the analysis of specialized, high-dimensional textual data. We integrate unsupervised methods such as Louvain, K-means, and Spectral clustering with doc2vec to enhance the detection of semantic patterns across a large corpus. As a case study, we apply this methodology to cybersecurity risk analysis using the MITRE ATT\&CK framework to structure and reduce the dimensionality of cyberattack tactics. Louvain clustering proved the most effective among the tested methods, achieving the best balance between cluster coherence and computational efficiency. Our approach identifies four \quotes{super tactics,} demonstrating how clustering improves thematic coherence and risk attribution. The results validate the utility of combining doc2vec with clustering, particularly Louvain, for enhancing topic modeling and text classification.
\end{abstract}

\medskip


\thispagestyle{empty}
\clearpage

\onehalfspacing
\setcounter{footnote}{0}
\renewcommand{\thefootnote}{\arabic{footnote}}
\setcounter{page}{1}

\section{Introduction}

Natural Language Processing (NLP) has become essential in analyzing large volumes of textual data, particularly in domains like cybersecurity, where identifying and interpreting risk-related information embedded in corporate documents is critical. Over the years, text classification techniques have evolved significantly from traditional dictionary-based approaches to more advanced vector-based models. Dictionary-based methods, once prevalent, rely on predefined lists of words or n-grams to classify texts according to specific themes or sentiments (\textit{e.g.}, positive/negative sentiment, or risk-related content). Pioneering studies such as those by \cite{AntweilerFrank2004,Garcia2013,JegadeeshWu2013,Arslan-AyaydinBoudtThewissen2016} use dictionary approaches to analyze the impact of sentiment in financial news and stock forums, showing their effectiveness in specific contexts. However, these methods are inherently limited, relying on static, human-curated dictionaries that may omit important nuances, especially in specialized fields like cybersecurity.

Recent advancements in NLP, mainly using word embeddings and vector-based models like word2vec and doc2vec, have addressed many of these limitations. By embedding words or entire paragraphs into vector space, these models capture semantic relationships between words, going beyond simple word frequency counts to consider word context and order. One of the critical advantages of vector-based models, such as those proposed by \cite{MikolovChenCorradoDean2013} and later extended in the paragraph-to-vector model by \cite{LeMikolov2014}, is their ability to understand the semantic similarity between paragraphs even when synonymous or domain-specific terms are used. This feature makes them particularly well-suited for specialized fields like cybersecurity, where language is highly technical and rapidly evolving. Applications word embeddings-related methods are beneficial for the automatic treatment of financial documents. For instance, \cite{SautnerLentVilkovZhang2023} and \cite{HassanHollandervLentTahoun2019} explore text classification to measure climate and political risks, respectively.

\cite{MaréchalMonnet2024W} leverage the doc2vec model and develop a method to attribute scores related to various contexts to textual documents. Since one context is almost always decomposed into several sub-contexts, they additionally create a method to reduce the sub-contexts' dimensionalities, with outputs of dimension size that the user can define. Their method was calibrated for cyber-risk scores attributed to firms-related documents (10-K reports) to test whether different cyber-risk scores command different risk-premia. We reproduce their aggregation method. First, the method embeds cybersecurity-related texts from the MITRE ATT\&CK knowledge base into vector representations, which they then use to analyze corporate financial disclosures to identify paragraphs that describe cyber risk. After clustering, the MITRE ATT\&CK database, structured around 14 cyberattack tactics and 785 sub-techniques, was reduced into four \quotes{supertactics}. We aim to provide an in-depth analysis of the resulting aggregation, showcasing a clear textual structure.

A key innovation in this approach is using clustering techniques to enhance the doc2vec model's ability to detect patterns across a sizeable textual corpus. Inspired by previous studies on topic modeling and clustering in NLP (\textit{e.g.}, \cite{CalomirisMamaysky2019, HassanHollandervLentTahoun2019}), we employ unsupervised clustering methods to group similar paragraphs, improving the identification of cybersecurity-related content. Specifically, integrating the Louvain clustering method, as outlined by \cite{BlondelGuillaumeLambiotteLefebvre2008}, allows us to partition paragraphs into clusters that reflect distinct aspects of cyber risk. This clustering technique is applied to the cybersecurity descriptions from MITRE ATT\&CK, enabling a more granular analysis of the semantic structure of the text.

The method implemented and studied in this paper has significant practical implications, allowing for efficient comparison between corporate risk disclosures and the highly technical cybersecurity descriptions in MITRE ATT\&CK. Building on the results of this method, \cite{CelenyMaréchal2023, MaréchalMonnet2024W}, calculated the cosine similarity between vector representations and quantitatively assess the degree to which specific sections of 10-K filings align with known cyberattack techniques, providing a robust mechanism for attributing cyber-risk scores to firms. Their additional work demonstrates the effectiveness of using doc2vec for thematic clustering and similarity measurement in large, heterogeneous textual datasets.

As stated before, our results reveal the existence of four critical super tactics: Preparation and Reconnaissance, Persistence and Evasion, Credential Movement, and Command and Data Manipulation. The Preparation and Reconnaissance super tactic encompasses tactics related to information gathering and preparatory actions before an attack. At the same time, Persistence and Evasion highlight techniques adversaries use to maintain access to a compromised network and avoid detection. The Credential Movement super tactic focuses on the theft and use of credentials for lateral movement within a network. Command and Data Manipulation involves controlling and manipulating compromised systems to achieve the attacker’s objectives.

Our comprehensive clustering analysis employs K-means, Louvain, and Spectral clustering methods to derive insights from the cosine similarity matrix constructed from MITRE ATT\&CK vectors. While the K-means method provides a foundational structure, it lacks exclusivity among super tactics, resulting in low balanced scores and high entropy. In contrast, the Louvain method improves the heterogeneity of clusters, particularly for tactics 5 (Resource Development) and 12 (Reconnaissance), while addressing issues related to overly similar paragraph structures by implementing similarity thresholds. Spectral clustering is a robust alternative, producing results comparable to Louvain's when hyperparameters are finely tuned. Ultimately, we select the Louvain method's output as our baseline for grouping tactics, balancing the need for coherent clusters without requiring excessive hyperparameter tuning. This comprehensive analysis contributes valuable insights into structuring textual data for practical application and contributes to the expanding literature on NLP-based analysis and clustering to reduce topic dimensionalities.


\section{Data and methodology}

\subsection{Cyberattacks knowledge database}
\label{section:cyber_risk_data}

\begin{figure}[H]
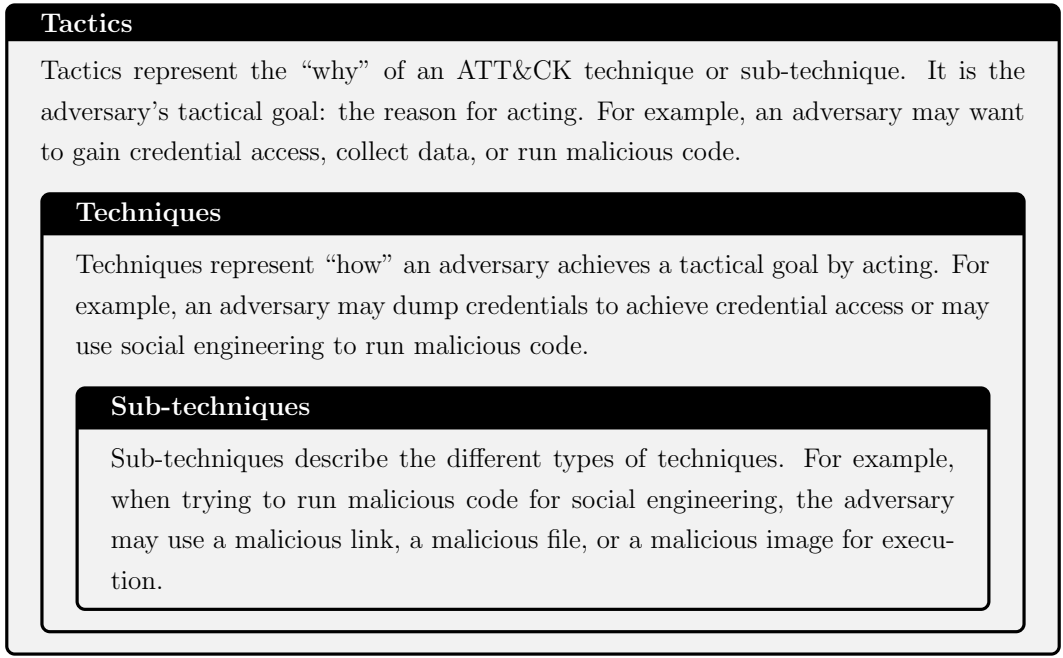

    \begin{adjustbox}{width=0.85\textwidth,center}
    \begin{tacticbox}
    Tactics represent the \quotes{why} of an ATT\&CK technique or sub-technique. It is the adversary's tactical goal: the reason for acting. For example, an adversary may want to gain credential access, collect data, or run malicious code. 
        \begin{idbox}
            Techniques represent \quotes{how} an adversary achieves a tactical goal by acting. For example, an adversary may dump credentials to achieve credential access or may use social engineering to run malicious code. 
            \begin{subidbox}
                Sub-techniques describe the different types of techniques. For example, when trying to run malicious code for social engineering, the adversary may use a malicious link, a malicious file, or a malicious image for execution.
            \end{subidbox}
        \end{idbox}
    \end{tacticbox}
    \end{adjustbox}
    \caption[\footnotesize Structure of MITRE ATT\&CK]{\centering \textbf{Structure of MITRE ATT\&CK}\bigskip
    }
    \label{fig:MITRE_structure}
\end{figure}

The MITRE ATT\&CK\footnote{\href{https://attack.mitre.org/}{https://attack.mitre.org/}} cybersecurity knowledge base is used as a reference for cyber attack descriptions. This knowledge base was created in 2013 to document cyber attack tactics, techniques, and procedures. It is structured by tactics, techniques, and sub-techniques as depicted in Figure \ref{fig:MITRE_structure}. There are 14 tactics: reconnaissance,  resource development, initial access, execution, persistence, privilege escalation, defense evasion, credential access, discovery, lateral movement, collection, command and control, exfiltration, and impact. There are 785 sub-techniques across all tactics. Two examples of sub-techniques are given in Table \ref{tab:MITRE_examples}.

\begin{table}[H]
\noindent\makebox[\textwidth]{
    \centering
    \begin{tabular}{l|c|l}
    \hline
    \hline
    & & \multicolumn{1}{c}{Description}\\
    \hline
    Tactic & Credential Access & \multirow{3}{70mm}{\small Adversaries may forge web cookies that can be used to gain access to web applications or Internet services. Web applications and services (hosted in cloud SaaS environments or on-premise servers) often use session cookies to authenticate and authorize user access.}\\[4.9ex]
    Technique & Forge Web Credentials & \\[4.9ex]
    Sub-technique & Web Cookies & \\[4.9ex]
    \hline
    Tactic & Reconnaissance & \multirow{3}{70mm}{\small Adversaries may gather employee names that can be used during targeting. Employee names can be used to derive email addresses and help guide other reconnaissance efforts and/or craft more believable lures. }\\[3ex]
    Technique &  Gather Victim Identity Information & \\[3ex]
    Sub-technique & Employee Names & \\[3ex]
    \hline
    \hline
    \end{tabular}
}
\caption[\footnotesize MITRE ATT\&CK sub-technique examples]{\centering \textbf{Examples of sub-techniques from MITRE ATT\&CK}\bigskip


}
\label{tab:MITRE_examples}
\end{table}

\newpage

\subsection{Textual and semantical training data}
The doc2vec method used in this paper was initially developed to identify cybersecurity-related paragraphs in US-listed companies' financial statements (10-K reports). We will briefly explain what a 10-K report is to clearly describe how the doc2vec model was initially designed and trained.

10-K statements are financial filings publicly traded companies submit annually to the U.S. Securities and Exchange Commission (SEC). They contain textual information such as companies' financial statements, risk factors, and executive compensation. The SEC’s Edgar archives index files were used to download and structure the 10-K reports. These files contain information about all the documents filed by firms for specific dates. From these files, a total of 64,988 10-K statements were identified.

\subsection{Methodology}
\label{section:methodology}

\subsubsection{Vector representation using doc2vec}

 We use the paragraph-to-vector model proposed by \cite{LeMikolov2014}, which is an extension of the word2vec model (\citealp{MikolovChenCorradoDean2013}). There are various advantages to working with this NLP approach compared to others, such as the dictionary approach. First, the comprehension of the method is semantical, meaning that it is not limited to a count of word frequencies. The word order impacts the resulting vector, and paragraphs with similar or synonym words will have close vector representations. Second, training the model with specific text that involves a particular vocabulary allows the incorporation of relatively unknown words. Finally, the resulting vectors have a dimension usually much smaller than vectors resulting from the dictionary approach. 

\begin{figure}[H] 
    \noindent\makebox[\textwidth]{%
    \includegraphics[scale=0.4]{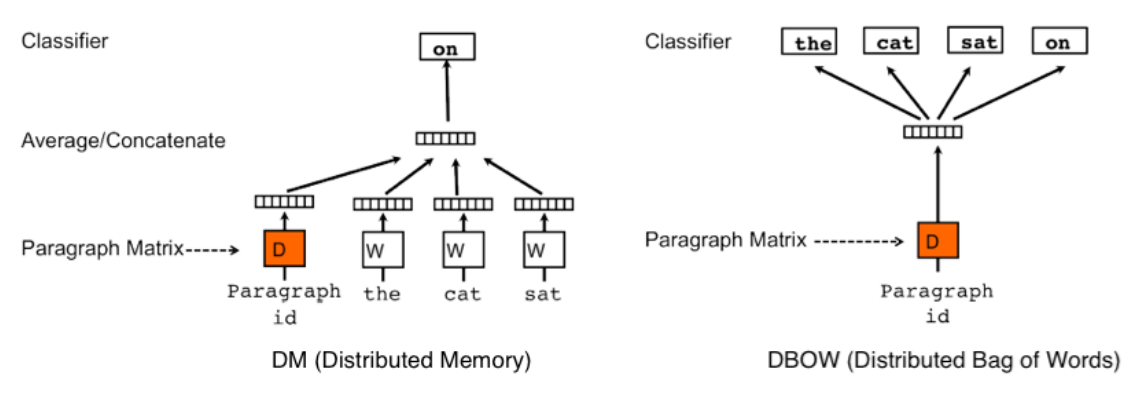}}
    
    \caption[\footnotesize Illustration of doc2vec training]{\textbf{Illustration of doc2vec training}}\bigskip
    \footnotesize{Illustration of the training of the neural network of the two versions of doc2vec, distributed memory model (DM) and distributed bag-of-words model (DBOW). The figure is taken from \cite{LeMikolov2014}.}
    \label{doc2vec_model}
\end{figure}

Two versions of the model exist: the distributed memory model (DM) and the distributed bag-of-words model (DBOW). In the DM, a neural network is trained as follows. First, a word is removed in a paragraph. Then, inputting the paragraph vector representation and context words (also in vector representation) surrounding the missing word, the neural network is optimized by trying to guess the missing word. In the DBOW, the neural network is trained to predict a series of words sampled from a paragraph using only the vector representation of the paragraph as input. Figure \ref{doc2vec_model} illustrates the training process of the two models.

\subsubsection{Cosine similarity}

Using the doc2vec method, all paragraphs of interest can be embedded into vectors. A common way to attribute a similarity score to two paragraph vectors is to take the cosine of the angle they form. Other ways exist, but only measuring the angle was proven more effective than considering the vectors' magnitude (see \citealp{AdosoglouLombardoPardalos2021}). This is because the latter is more affected by the random initialization of weights during training in the neural network that outputs the vectors.

\subsubsection{Training of doc2vec}

The training data and details, the hyperparameters and their validation, and the final model choice are extensively covered in \cite{CelenyMaréchal2023}. This work uses their saved doc2vec model.\footnote{ \url{https://github.com/technometrics-lab/17-Cyber-risk_and_the_cross-section_of_stock_returns}}. We report some relevant details in this section.

In \cite{CelenyMaréchal2023}, they use the implementation by Gensim called doc2vec\footnote{\url{https://radimrehurek.com/gensim/models/doc2vec.html}}. To train the model, they use the paragraphs from 10-K statements filed in 2007 and the 785 sub-technique descriptions from MITRE ATT\&CK, which amount to more than 1.7 million training paragraphs. Utilizing this training sample, they train DM and DBOW doc2vec models with various vector dimensions, epochs, and window sizes. The baseline for the hyperparameters is taken from \cite{LauBaldwin2016} (see Table \ref{tab:doc2vec_parameters}).

To select the best model, they first compute vector representations of paragraphs from 10 randomly chosen 10-K statements from 2008 (the validation sample) and vector representations of the paragraphs from MITRE ATT\&CK, using various trained doc2vec models. For each 10-K paragraph, they identify the highest cosine similarity score with any of the MITRE ATT\&CK paragraph vectors. The model that produces the highest proportion of 10-K paragraphs with strong similarity scores to MITRE ATT\&CK paragraphs is deemed the best. In other words, they select the model identifying the most significant proportion of cybersecurity-related paragraphs in the 10-Ks. Table \ref{tab:doc2vec_parameters} presents the parameters of the best-performing doc2vec model, which is used for the remainder of this study. Table \ref{tab:top_paragraphs_doc2vec_validation} presents the top-scoring paragraphs from the validation sample (after pre-processing). 

\begin{table}[h]
    \resizebox{\textwidth}{!}{
    \begin{tabular}{lccccccc}
        \textbf{Method} & \textbf{Training Size} & \textbf{Vector Size} & \textbf{Window Size} & \textbf{Min Count} & \textbf{Sub-Sampling} & \textbf{Negative Sampling} & \textbf{Epoch}\\  \cmidrule(lr){1-1} \cmidrule(lr){2-2} \cmidrule(lr){3-3} \cmidrule(lr){4-4} \cmidrule(lr){5-5} \cmidrule(lr){6-6} \cmidrule(lr){7-7} \cmidrule(lr){8-8} 
        DBOW & 1.7M & 200 & 15 & 5 & $10^{-5}$ & 5 & 50 \\
    \end{tabular}}
    \caption[\footnotesize doc2vec parameters]{\textbf{doc2vec parameters}}\bigskip
    \footnotesize{Parameters of the chosen doc2vec model. DBOW stands for distributed bag-of-words.}
    
    \label{tab:doc2vec_parameters}
\end{table}
\newpage

\begin{table}[H]
\vspace{-3cm}
\noindent\makebox[\textwidth]{%
    \begin{tabular}{cp{90mm}cc}
        \toprule
        \toprule
        Score & \multicolumn{1}{c}{Preprocessed paragraph} & Firm ID (Ticker) & Tactic\\
        \hline
        0.593 & \footnotesize currently available internet browsers allow users modify browser settings remove cookies prevent cookies stored hard drives however third persons able penetrate network security gain access otherwise misappropriate users personal information subject liability liability include claims misuses personal information unauthorized marketing purposes unauthorized use credit cards & VSTY & Defense Evasion \\
        0.590 & \footnotesize network security data recovery measures may adequate protect computer viruses break ins similar disruptions unauthorized tampering computer systems theft sabotage type security breach respect proprietary confidential information electronically stored including research clinical data material adverse impact business operating results financial condition & LXRX &
        Collection \\
        0.583 & \footnotesize domain names derive value individual ability remember names therefore assurance domain name lose value example users begin rely mechanisms domain names access online resources government regulation internet regulation increasing number laws regulations pertaining internet & VSTY & Credential Access\\
        0.577 & \footnotesize perceived actual unauthorized disclosure information collect breach security harm business factors beyond control cause interruptions operations may adversely affect reputation marketplace business financial condition results operations timely development implementation continuous uninterrupted performance hardware network applications internet systems including may provided third parties important facets delivery products services customers & MDAS & Credential Access\\
        0.571 & \footnotesize unauthorized parties may attempt copy aspects products obtain use information regard proprietary others may independently develop otherwise acquire similar competing technologies methods design around patents cases rely trade secret laws confidentiality agreements protect confidential proprietary information processes technology & CSCD & Collection\\
        \bottomrule
        \bottomrule
    \end{tabular}  
}
\caption[\footnotesize Top scoring paragraphs from the doc2vec validation sample]{\textbf{Top scoring paragraphs from the doc2vec validation sample}}\bigskip
\footnotesize{The paragraphs are shown after preprocessing. Tactic refers to the MITRE ATT\&CK tactic the paragraph is most similar to, as measured by cosine similarity. The table is taken from \cite{CelenyMaréchal2023}.}
\label{tab:top_paragraphs_doc2vec_validation}
\end{table}

\begin{table}[H]
\vspace{-3cm}
\noindent\makebox[\textwidth]{%
    \begin{tabular}{cp{90mm}cc}
        \toprule
        \toprule
        Score & \multicolumn{1}{c}{Preprocessed paragraph} & Firm ID (Ticker) & Tactic\\
        \hline
        0.570 & \footnotesize possible cookies may become subject laws limiting prohibiting use term cookies refers information keyed specific server file pathway directory location stored user hard drive possibly without user knowledge used among things track demographic information target advertising & VSTY & Discovery\\
        0.561 & \footnotesize cannot certain advances computer capabilities discoveries field cryptography developments result compromise breach algorithms use protect content transactions website proprietary information databases anyone able circumvent security measures misappropriate proprietary confidential customer company information cause interruptions operations & VSTY & Impact\\
        0.558 & \footnotesize ordering delivery customers ready place order proceed shopping cart function directly checkout page orders placed online website via toll free telephone number customer service agents available take orders customers access internet uncomfortable placing order online & VSTY & Credential Access\\
        0.557 & \footnotesize process allows identify catalogue embryonic stem cell clone dna sequence trapped gene select embryonic stem cell clones dna sequence generation knockout mice used gene trapping technology automated process create omnibank library frozen gene knockout embryonic stem cell clones identified dna sequence relational database & LXRX & Persistence \\
        0.556 & \footnotesize believe systematic biology driven approach technology platform makes possible provide substantial advantages alternative approaches drug target discovery particular believe comprehensive nature approach allows uncover potential drug targets within context mammalian physiology might missed narrowly focused efforts & LXRX & Discovery\\
        0.554 & \footnotesize concerns security internet may reduce use website impede growth significant barrier confidential communications internet need security rely ssl encryption technology designed prevent customer credit card data transaction process current credit card practices merchant liable fraudulent credit card transactions case transactions process merchant obtain cardholder signature & VSTY & Credential Access\\
        \bottomrule
        \bottomrule
    \end{tabular}
}
\caption*{Table \ref*{tab:top_paragraphs_doc2vec_validation}: \textbf{Top scoring paragraphs from the doc2vec validation sample (continued)}}
\end{table}

\newpage

\subsubsection{Clustering methods}


A natural way of splitting the cybersecurity textual database into different categories comes from the written structure of MITRE ATT\&CK, with 14 categories already established as tactics. In \cite{MaréchalMonnet2024W}, the similarity of each tactic with 10-K paragraphs resulted in 14 different cyberscores which was too high to maintain their explanatory power. Therefore, they argue in favor of aggregating the 14 tactics into a few super tactics to mitigate this effect. We aim to provide a detailed account of this approach to clustering textual data and will, therefore, follow a similar methodology.

Using the doc2vec method and the cosine similarity score, We can transform every 785 sub-techniques (paragraphs) of MITRE ATT\&CK into vectors and compare their similarity. This process results in a similarity matrix of dimension 785 by 785, onto which clustering methods can be applied. Indeed, the similarity matrix can be understood as the representation of a network where every 785 nodes (paragraphs) are connected by edge values weighted by their similarity. In this context, We present three classical clustering methods.

The first and most simplistic clustering method is K-Means. Note that since the input similarity matrix is based on cosine similarity, it is instead designated as spherical K-Means, where the distance between each point to class into K categories is understood as the angle between the vectors defined by those points rather than the Euclidian distance between those points. Either version of K-Means works as follows: It begins by randomly setting initial cluster centroids, then iteratively assigns each data point (paragraphs) to the nearest centroid and updates the centroids by recalculating their mean positions among their associated data points. The process is repeated until convergence. Note that although the K-Means algorithm always converges, it is relatively dependent on the initial centroid guess. The user must choose the number of clusters K without prior knowledge. The algorithm generally produces rough results but often reveals an initial simple structure in the similarity of the provided data.

The second method is much more potent as it requires no prior hyperparameters; thus, the number of clusters is an output of the method. The Louvain method provides a straightforward way to identify clusters (groups of nodes within a graph that are more densely connected) in a network (see, \textit{e.g.}, \citealp{CalomirisMamaysky2019}). To explain the Louvain method, We first need to introduce the notion of modularity. It is defined as a value in the $[-1/2, 1]$ that measures the density of links within communities compared to links between. For a weighted graph, modularity is defined as:

\[
Q = \frac{1}{2m} \sum_{i=1}^{N} \sum_{j=1}^{N} \left[ S_{ij} - \frac{k_ik_j}{2m} \right] \delta(c_i, c_j),
\]

where $S_{ij}$ represents the edge weight between nodes $i$ and $j$, in this case, this is the similarity matrix. $k_i$ and $k_j$ are the sum of the weights of the edges attached to nodes $i$ and $j$, respectively. $m$ is the sum of all the graph's edge weights. $N$ is the total number of nodes in the graph.  $c_i$ and $c_j$ are the communities to which the nodes $i$ and $j$ belong and $\delta$ is the Kronecker delta function.
The Louvain method works as follows. Initially, each node is assigned to its community. Then, the method iterates through two phases: the first phase optimizes modularity locally by moving individual nodes between communities to maximize the increase in modularity. The second phase aggregates the nodes in each community in the first phase into single nodes and builds a new network, where the communities found in the first phase are treated as nodes. Phases one and two are repeated until no further improvement in modularity is possible. The final partitioning of nodes into communities is returned as a result.

The third clustering method is spherical K-means on a dimensionally reduced similarity matrix. The spectral clustering method works as follows. First, the degree matrix $D$ is constructed. it consists in a diagonal matrix where each entry $D_{ii}$ represents the sum of similarities for node $i$ and is computed as $D_{ii} = \sum_{j} S_{ij}$. The Laplacian matrix $L$ is defined as $L=D-S$. The spectral clustering algorithm computes the eigenvectors and eigenvalues of the Laplacian matrix $L$. Let $\lambda_1, \lambda_2, \ldots, \lambda_N$ be the eigenvalues and $v_1, v_2, \ldots, v_N$ be the corresponding eigenvectors. After obtaining the eigenvectors, We select the $K$ eigenvectors corresponding to the $K$ smallest eigenvalues (excluding the smallest eigenvalue, typically zero). We arrange these eigenvectors as columns in a matrix $V$ of dimension K by N. Finally, We perform clustering on the rows of the matrix $V$ using the k-means clustering method. The power of this approach is that We can choose the number of features necessary to perform a satisfying clustering (going from N=785 to K<20, for example, can radically improve the clustering by getting rid of superfluous dimensions).

Finally, a form of scoring is needed to find the best clustering output produced by the wide range of hyper-parameters and choices of method. In this paper, We propose a relatively simple but efficient approach that requires initial labelalization of each node. Each paragraph (node) is a sub-technique belonging to one of the 14 tactics of MITRE ATT\&CK. Thus, they naturally belong exclusively to 14 \quotes{sub-clusters.} The discrimination of clustering methods works on the following two requirements.

First, We want the paragraphs belonging to one tactic (sub-cluster) to belong to the same super-tactic (to the same cluster found by the method). Indeed, the paragraphs are initially classed by the creator of MITRE ATT\&CK together because they share common characteristics. It would not be very sensible to spread them across different super-tactics (clusters) once the clustering method is applied. Thus, a measure of sub-cluster heterogeneity among clusters is needed. We use Shannon entropy, defined as follows:
\begin{equation}
    H_{sub_j}=-\sum_{i=1}^{nb. clusters} P(sub_j)_i\log P(sub_j)_i
\end{equation}
Where $P(sub_j)_i$ is the proportion of paragraphs of sub-cluster (tactic) j belonging to cluster (super tactic) $i$. Intuitively, if we are in an ideal case and the paragraphs of a sub-cluster $j$ are entirely contained in cluster 1 we would have $P(sub_j)_1=1$ and $P(sub_j)_i=0$ for $i\neq1 $, thus leading to $H_{sub_j}=0$ being minimal (mind the minus sign in the equation and the logarithm on number lower than 1). If we start to spread the paragraph of the sub-cluster among other clusters, the $P(sub_j)_i$
become different from 0 and 1 and $H_{sub_j}$ gradually increases. To reduce the 14 $H_{sub_j}$ to one score of discrimination, We sum them all, thus obtaining the Entropy sum, the sub-cluster heterogeneity among clusters. The heterogeneity is high when the Entropy sum is low.

Second, We need a score to counter the following extreme case. All sub-clusters, but one may be classed into one cluster and the last sub-cluster into a second cluster. This would lead to a minimum Entropy sum of 0 but would have no value for our application. We want the sub-cluster to be reasonably spread out among the cluster. To translate this idea into a meaningful score, we create a Balanced score, defined as the standard deviation of the label counts. In other words, the clustering method produced an ordered list of 785 values corresponding to the label of the cluster each paragraph belongs to. For each label, We count the number of occurrences on the list. If the paragraphs are relatively well spread out across the cluster, then taking the standard deviation of all the count of the labels should be low since each cluster would contain approximately the same number of paragraphs. The last case to worry about is that the balanced score could be low, but the paragraph would be randomly spread across the cluster, thus not reflecting the initial structure of MITRE ATT\&CK tactics (sub-clusters). To counter that, it is sufficient to consider the Entropy sum. 

Considering the method that outputs the lowest Entropy sum and the lowest balanced score, We can effectively discriminate the different clustering methods' outputs. Note that there is no guideline regarding the optimal trade-off between the two scores, \textit{i.e.} what additional amount is optimal to forfeit to the Entropy sum to lower the Balanced score and inversely. 

Finally, the whole clustering process described here must be seen more as a guideline tool. Indeed, after choosing the best method, We class each paragraph in the cluster where most of its sub-cluster belongs, regardless of the method's output for the misplaced paragraphs. The structure of MITRE ATT\&CK is probably more coherent than the output of any unsupervised clustering method. However, the new clustering structure output based on the cosine similarity matrix could maximize the likelihood of reducing the correlation between the different sub-cyber scores based on cosine similarities.

\newpage

\section{Results}

\subsection{Clustering of MITRE ATT\&CK}

\begin{landscape}
 \begin{figure}[H] 
 \centering
    \noindent\makebox[\textwidth]{%
    \centering
    \includegraphics[width=1.4\textwidth,keepaspectratio]{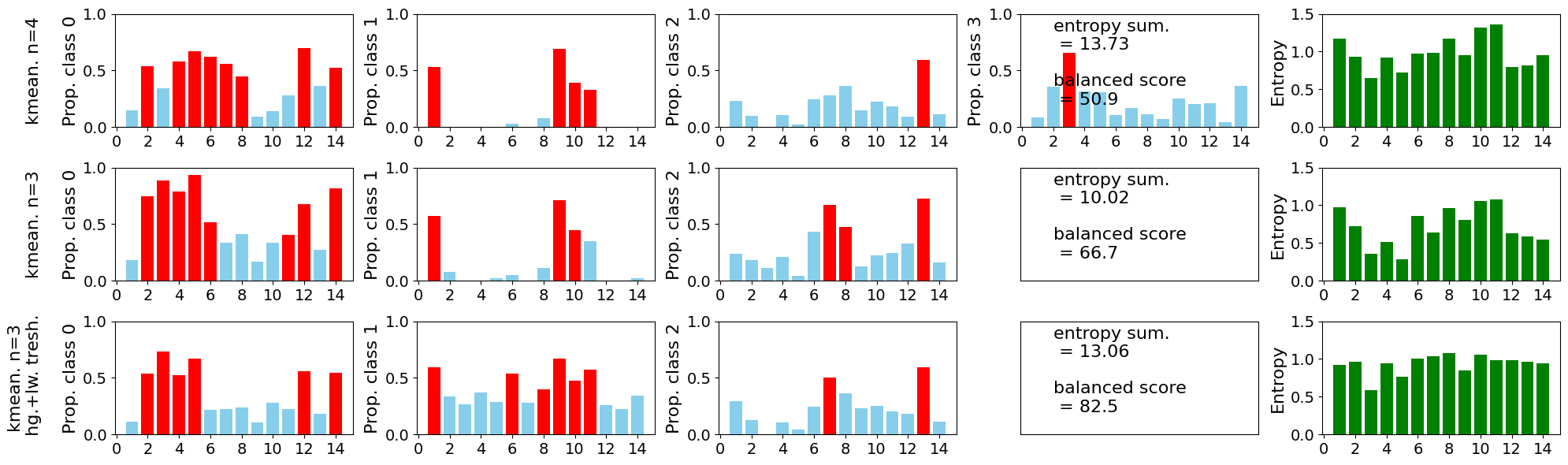}
    }
    \caption[\footnotesize Clustering results part.1]{\textbf{Clustering results part.1}}\label{clustering1}\bigskip
    \footnotesize{
    \begin{flushleft}
    This figure presents the results of each clustering method indicated on the left. The figure in red and blue represents $P(sub_j)_i$, the proportion of paragraphs of sub-cluster (tactic) j belonging to cluster (super tactic) i. The 14 sub-cluster labels are on the x-axis of each figure, and the cluster labels correspond to the columns (class 0 to 3, here). If the proportion is in red, it means it is the highest in the cluster (in other clusters/columns, the same sub-cluster will be in blue). We also report the entropy sum and the balanced score on the figure for each method. Finally, the individual Shannon entropy of each sub-cluster is reported in green in the last column. In the name of the method, we also indicate the hyperparameters of the method. Here, $n$ corresponds to the number of clusters imposed by the k-means method. \quotes{hg. tresh.} and \quotes{lw. tresh.} corresponds to a change applied to the similarity matrix. If the value in the similarity matrix is lower than 0.25, it is changed to 0 (lower threshold), and if the similarity is higher than 0.85, it is changed to 0.5 (higher threshold). In part.2 and part.3 \quotes{egn} corresponds to the $K$ eigenvectors in the spectral clustering. We also made the output clusters of each method match. Hence, the comparison is simpler (otherwise, what the Louvain method called cluster 2 is not necessarily cluster 2 for the k-means method). The following list shows the corresponding number of each tactic : 1: Persistence, 2: Command and Control, 3: Impact, 4: Initial Access, 5: Resource Development, 6: Collection, 7: Exfiltration, 8: Credential Access, 9: Privilege Escalation, 10: Execution, 11: Defense Evasion, 12: Reconnaissance, 13: Lateral Movement, 14: Discovery.
    \end{flushleft}} 
\end{figure}
\end{landscape}

\begin{landscape}
 \begin{figure}[H] 
 \centering
    \noindent\makebox[\textwidth]{%
    \centering
    \includegraphics[width=1.4\textwidth,keepaspectratio]{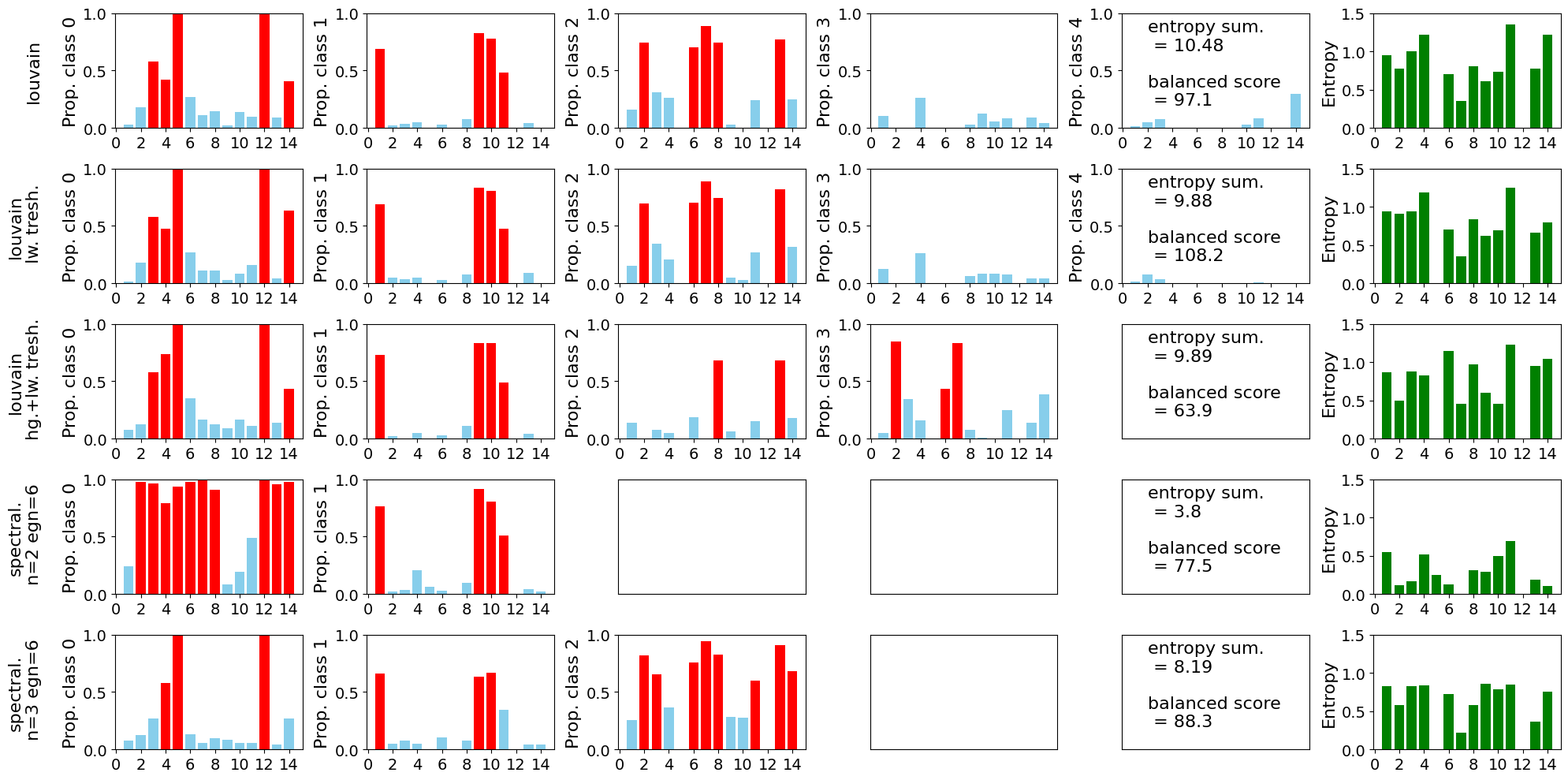}
    }
    \caption[\footnotesize Clustering results part.2]{\textbf{Clustering results part.2}}\label{clustering2}\bigskip
    \footnotesize{
    \begin{flushleft}
    \end{flushleft}} 
\end{figure}
\end{landscape}

We apply the clustering methods on the cosine similarity matrix created from MITRE ATT\&CK paragraphs vector embeddings. This allows for identifying the relevant paragraphs tied to previously mentioned super tactics (command and data manipulation, credential movement, persistence and evasion, and preparation and reconnaissance). We report the results of various attempts with different clustering methods in Figures \ref{clustering1}, \ref{clustering2}, and \ref{clustering3}. The K-means methods provide a coherent but crude initial structure we report in Figure \ref{clustering1}. Indeed, the paragraphs tend to be well spread across the super tactics (clusters) but at the cost of heterogeneity, with the exclusivity of a tactic in a super tactic being inexistent. This results in a low balanced score at the cost of entropy, as depicted in Figure \ref{clustering_scores}. 

Figure \ref{clustering2} shows the performance of the Louvain method. This method dramatically improves the heterogeneity, especially with tactics 5 and 12 (resource development and reconnaissance) exclusive to cluster 1 (the super tactic: preparation and reconnaissance). However, not putting a threshold on the inputted similarity matrix component induces the Louvain method to create two superfluous clusters. Hence, we include those restrictions. Indeed, when comparing two paragraphs of MITRE ATT\&CK, it is not uncommon to encounter sentences with similar structures for different semantic content. Thus, we tone down the similarity of a highly too similar paragraph with a higher threshold. Conversely, we define a lower threshold such that similarities that are too low and, therefore, most likely noise that reflects no similarity are set to zero. 

The last method can be seen as a safeguard for the output of the Louvain method. Applying the spectral clustering method, we retrieve the structure previously encountered with higher heterogeneity than with K-means. If the hyperparameters are correctly tuned, the output is similar to the Louvain method's, particularly for $n=4$ and $egn=6$. We report the results in Figure \ref{clustering3}. Including more dimensions (higher $egn$ value) adds noise and decreases the clustering quality.

Finally, we select the output of the Louvain method as a baseline to group the tactics without splitting them across super tactics. Although Figure \ref{clustering_scores} shows that outputs of other methods may be slightly better, we favor the Louvain method since no additional hyperparameters tuning is required.

\newpage
\thispagestyle{empty}

\thispagestyle{empty}

\begin{landscape}
 \begin{figure}[H]
 \centering
    \noindent\makebox[\textwidth]{%
    \centering
    \includegraphics[width=1.4\textwidth,keepaspectratio]{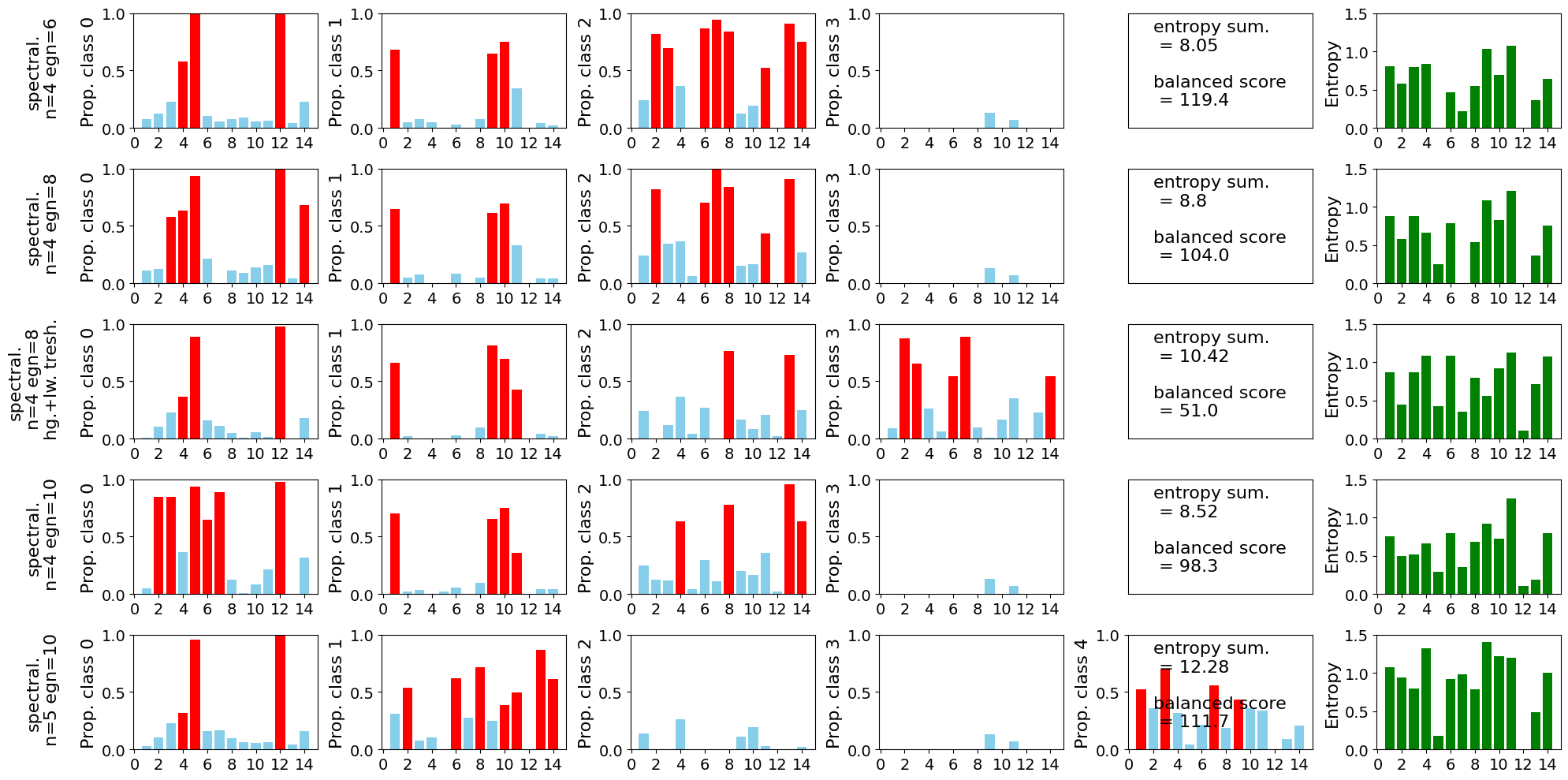}
    }
    \caption[\footnotesize Clustering results part.3]{\textbf{Clustering results part.3}}\label{clustering3}\bigskip
    \footnotesize{
    \begin{flushleft}
    
    \end{flushleft}} 
\end{figure}
\thispagestyle{empty}
\end{landscape}

\begin{figure}[H] 
    \noindent\makebox[\textwidth]{%
    \includegraphics[width=0.9\textwidth,keepaspectratio]{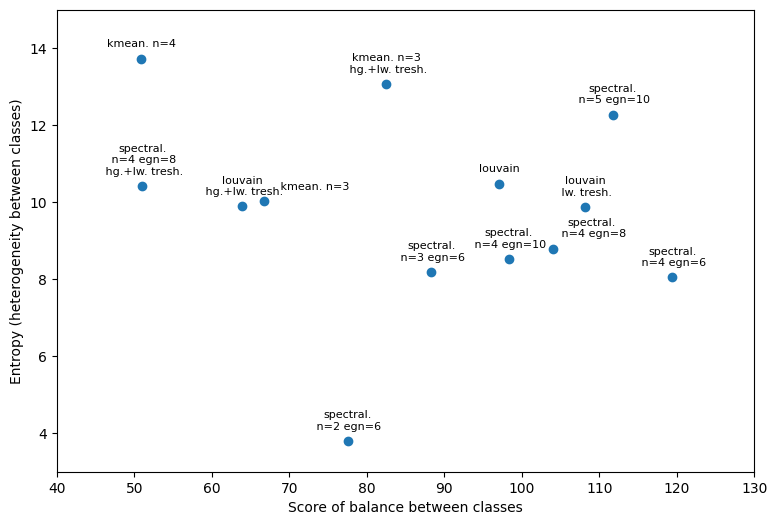}
    }
    \caption[\footnotesize Comparison of clustering scores: Entropy sum and Balanced score]{\textbf{Comparison of clustering scores: Entropy sum and Balanced score}}\label{clustering_scores}\bigskip
    \footnotesize{Each clustering method of Figure \ref{clustering1}, \ref{clustering2}, and \ref{clustering3} is presented here using their respective entropy sum and balanced score. Recall that the aim was to reduce both scores to distinguish the best clustering method. Also, note that there is no guideline regarding what additional amount
is optimal to forfeit to the entropy sum to lower the balanced score and inversely. 
}
\end{figure}

\begin{figure}[H] 
    \noindent\makebox[\textwidth]{%
    \includegraphics[width=0.9\textwidth,keepaspectratio]{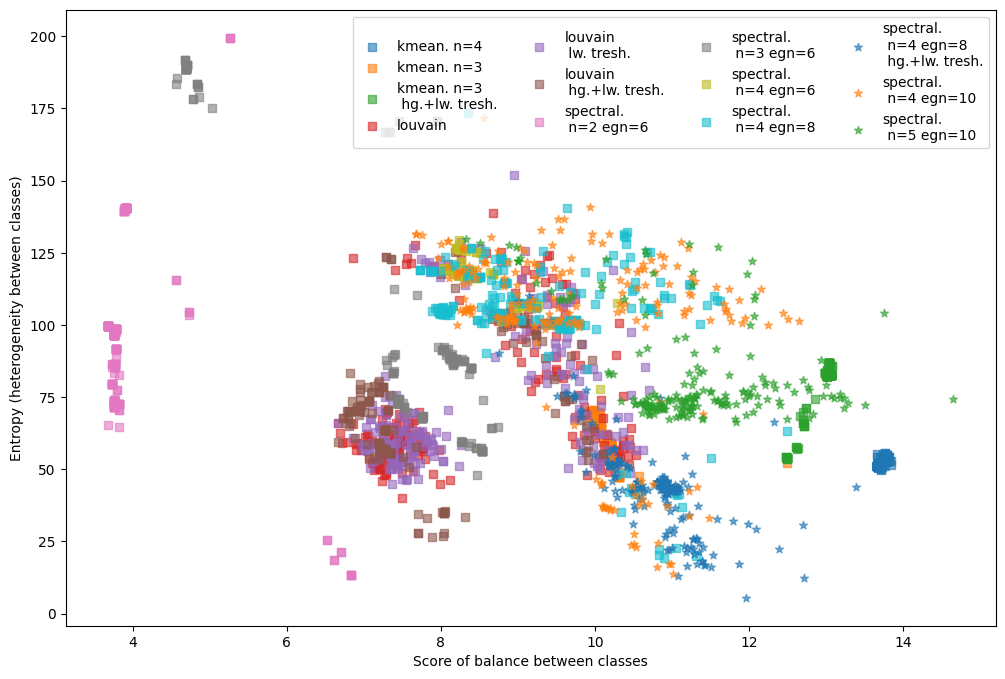}
    }
    \caption[\footnotesize Clustering scores: Entropy sum and Balanced score N=200]{\textbf{Clustering scores: Entropy sum and Balanced score N=200}}\label{200}\bigskip
    \footnotesize{The experiment of figure \ref{clustering_scores} is reproduced here by reapplying 200 times the involved clustering methods to avoid the bias of random initialization. Each clustering method is displayed using their respective entropy sum and balanced score.  
}
\end{figure}

\newpage 

\noindent This yields the following super tactics, named after their content: 

\textbf{Preparation and Reconnaissance}: 
This super tactic encompasses adversaries' tactics to prepare and gather information before launching an attack. \textbf{Impact} involves actions that disrupt, destroy, or manipulate systems and data to achieve the attacker’s objectives. \textbf{Initial Access} includes techniques adversaries use to gain an initial foothold within a network, such as exploiting vulnerabilities or spear phishing. \textbf{Resource Development} entails the acquisition of resources like infrastructure, tools, and credentials necessary to support operations. \textbf{Reconnaissance} involves gathering information about the target environment to identify potential entry points and vulnerabilities. \textbf{Discovery} refers to techniques used to explore and map the target environment, such as network scanning and enumeration.\newline

\textbf{Persistence and Evasion}: 
Once inside a target network, adversaries employ these tactics to maintain their foothold and avoid detection. \textbf{Persistence} ensures the attacker can maintain access even if the system is rebooted or credentials are changed by installing malware or creating rogue accounts. \textbf{Privilege Escalation} involves gaining higher-level permissions to access more sensitive information and critical systems. \textbf{Execution} refers to running malicious code on a victim system, often necessary to carry out the attacker’s objectives. \textbf{Defense Evasion} includes a variety of methods to avoid detection and thwart defensive measures, such as disabling security software, obfuscating code, or using fileless malware.\newline

\textbf{Credential Movement}: 
This group focuses on techniques to steal and use credentials to move within a network. \textbf{Credential Access} involves obtaining account names, passwords, and other secrets that allow attackers to authenticate themselves as legitimate users. Techniques include keylogging, credential dumping, and brute force attacks. \textbf{Lateral Movement} is moving through a network to find and access additional targets or more valuable data. This can be done using remote services, exploiting trust relationships, or leveraging legitimate credentials to access other systems and resources.\newline

\textbf{Command and Data Manipulation}: 
In this phase, adversaries control compromised systems and manipulate data to achieve their goals. \textbf{Command and Control} involves establishing a communication channel with the compromised environment to issue commands and control malware. This can be achieved through web traffic, DNS, or custom communication protocols with command servers. \textbf{Collection} refers to gathering sensitive information from compromised systems, such as capturing screenshots, logging keystrokes, or accessing stored files. \textbf{Exfiltration} involves transferring the collected data from the target network to an external location controlled by the adversary, often using encrypted channels or covert methods to avoid detection.\newline

\newpage

\section{Conclusion}

This study presents an innovative and rigorous approach to aggregate textual data already categorized in relevant sections/chapters. We used the MITRE ATT\&CK framework as a practical example and processed the textual database with an advanced Natural Language Processing (NLP) technique, specifically the doc2vec model. Our primary objective is maintaining a textual structure after aggregating the different sections using a simple but efficient double scores system.

Our clustering analysis identified four critical super tactics (aggregation of tactics): Preparation and Reconnaissance, Persistence and Evasion, Credential Movement, and Command and Data Manipulation. Each super tactic encapsulates a range of specific techniques employed by adversaries. Preparation and Reconnaissance encompasses tactics such as initial access and resource development, focusing on how adversaries gather information and prepare for attacks. Persistence and Evasion include techniques that ensure attackers maintain access to compromised systems while avoiding detection, highlighting the importance of stealth in cyber operations. Credential Movement addresses methods attackers use to gain and exploit credentials, facilitating lateral movement within networks to access sensitive information. Command and Data Manipulation illustrates how adversaries control compromised systems and exfiltrate data, showcasing the ultimate goals of cyber intrusion. We observe that each super-tactic is a natural and causal step from the previous super-tactic, further highlighting the coherence of the aggregated structure. 

We explore multiple methods to refine our clustering results, including K-means, Louvain, and Spectral clustering. Our analysis revealed that while K-means provided a coherent but somewhat crude initial structure, the Louvain method significantly enhanced heterogeneity, particularly in distinguishing tactics like Resource Development and Reconnaissance. However, we also identified the necessity of fine-tuning similarity thresholds to avoid the creation of superfluous clusters due to syntactically similar but semantically distinct paragraphs. The Louvain method was chosen as our baseline clustering approach, not because of its slight performance advantages over other methods but because of its practicality. This method yields meaningful clusters without extensive hyperparameter tuning, making it a straightforward and practical choice. 

While we use a case study focused on cybersecurity risks using the MITRE ATT\&CK framework, our methodology has much broader potential. The combination of doc2vec embeddings with clustering techniques, particularly the Louvain method, can be readily adapted to analyze other forms of risk, such as legislative, geopolitical, or environmental risks. This approach could efficiently identify and categorize risk factors embedded in corporate or regulatory texts by substituting the MITRE ATT\&CK knowledge base with a similarly structured dataset relevant to these domains. Maintaining a coherent, structured aggregation of textual information makes this method a powerful tool for exploring risk in various sectors, such as finance or public policy.

\clearpage

\bibliographystyle{styles/jfe}
\bibliography{bibliography.bib}

\clearpage
\section*{Appendix}

\setcounter{table}{0}
\renewcommand{\thetable}{A\arabic{table}}
\setcounter{figure}{0}
\renewcommand{\thefigure}{A\arabic{figure}}

\begin{figure}[H] 
    \noindent\makebox[\textwidth]{%
    \includegraphics[width=0.9\textwidth,keepaspectratio]{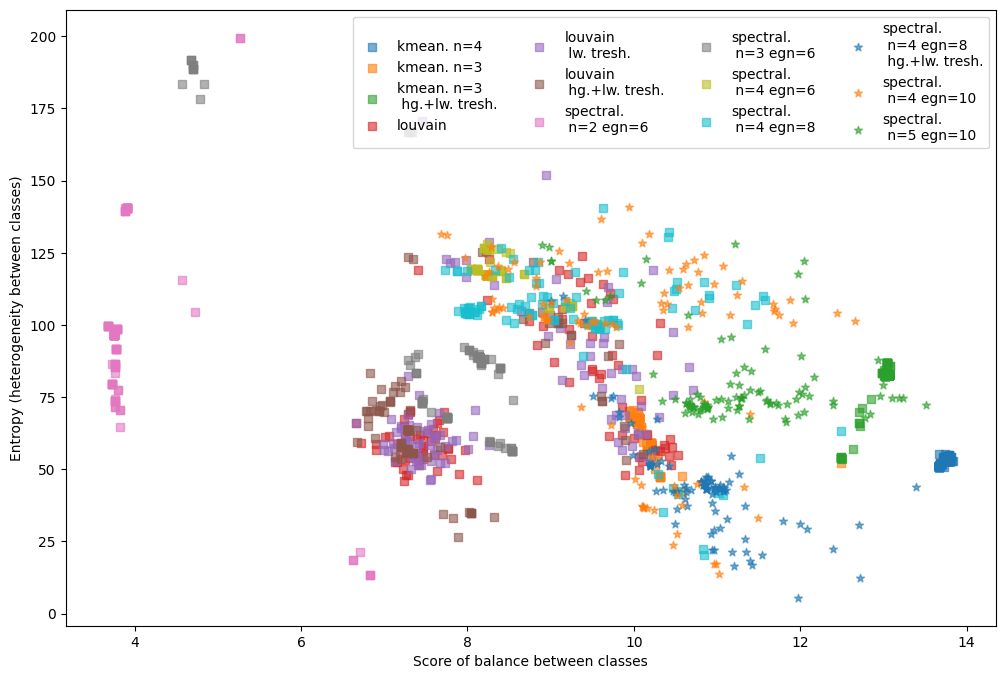}
    }
    \caption[\footnotesize Clustering scores: Entropy sum and Balanced score N=100]{\textbf{Clustering scores: Entropy sum and Balanced score N=100}}\label{100}\bigskip
    \footnotesize{The experiment of figure \ref{clustering_scores} is reproduced here by reapplying 100 times the involved clustering methods to avoid the bias of random initialization. Each clustering method is displayed using their respective entropy sum and balanced score.  
}
\end{figure}

\begin{figure}[H] 
    \noindent\makebox[\textwidth]{%
    \includegraphics[width=0.9\textwidth,keepaspectratio]{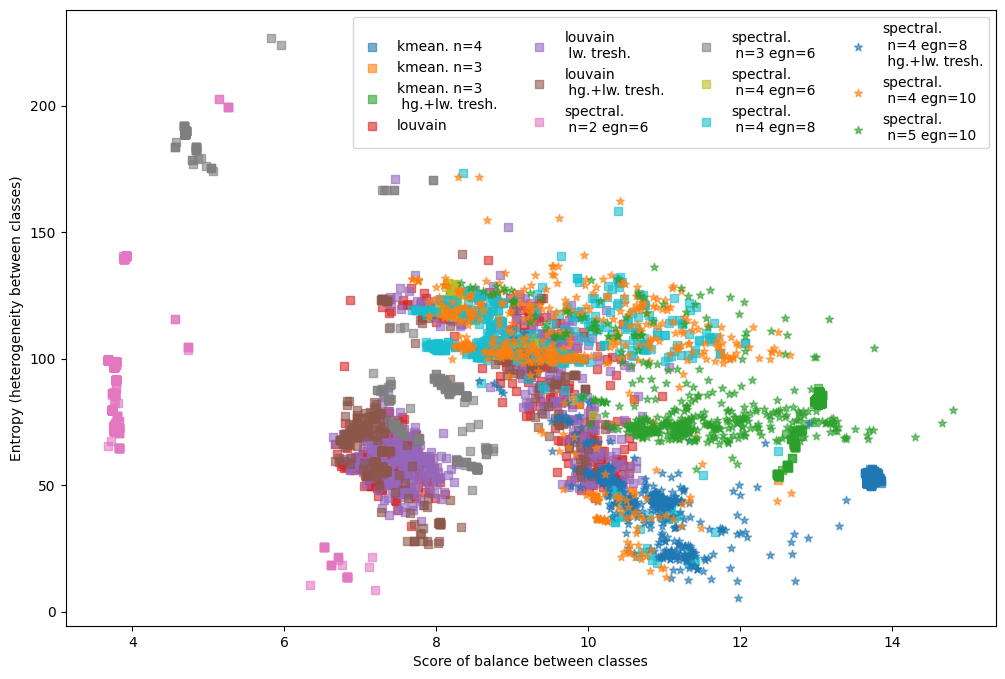}
    }
    \caption[\footnotesize Clustering scores: Entropy sum and Balanced score N=500]{\textbf{Clustering scores: Entropy sum and Balanced score N=500}}\label{500}\bigskip
    \footnotesize{The experiment of figure \ref{clustering_scores} is reproduced here by reapplying 500 times the involved clustering methods to avoid the bias of random initialization. Each clustering method is displayed using their respective entropy sum and balanced score.  
}
\end{figure}

\begin{figure}[H] 
    \noindent\makebox[\textwidth]{%
    \includegraphics[width=0.9\textwidth,keepaspectratio]{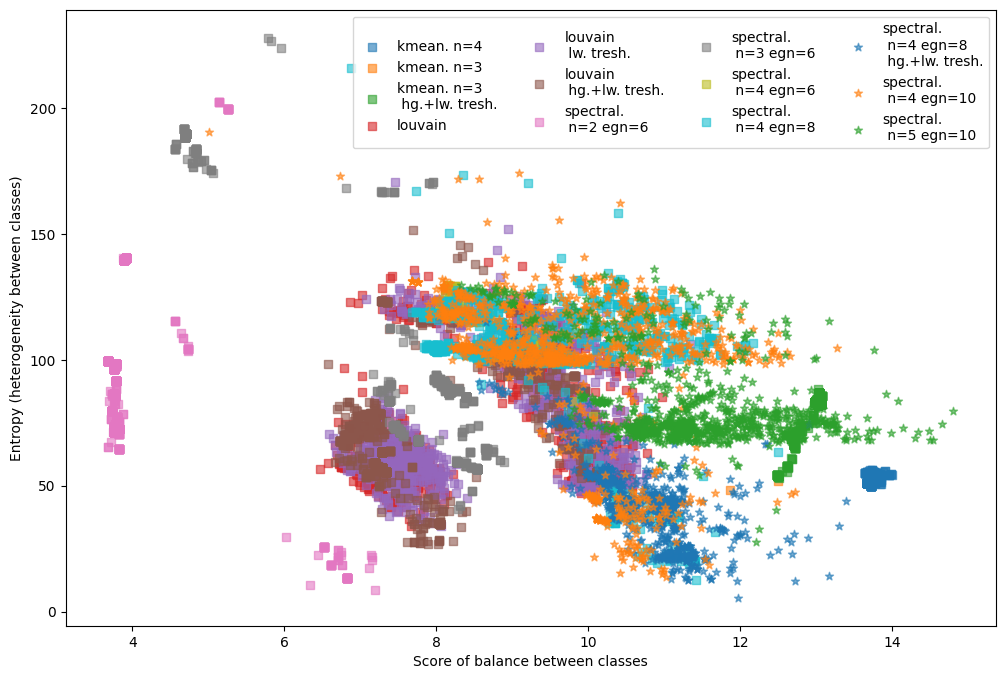}
    }
    \caption[\footnotesize Clustering scores: Entropy sum and Balanced score N=1000]{\textbf{Clustering scores: Entropy sum and Balanced score N=1000}}\label{1000}\bigskip
    \footnotesize{The experiment of figure \ref{clustering_scores} is reproduced here by reapplying 1000 times the involved clustering methods to avoid the bias of random initialization. Each clustering method is displayed using their respective entropy sum and balanced score.  
}
\end{figure}

\end{document}